\documentclass[aps,prl,twocolumn,showpacs,groupedaddress]{revtex4}  
\usepackage{graphicx,subfigure}  
\usepackage{dcolumn}   
\usepackage{bm}        
\usepackage{amssymb}   

\usepackage{pifont}    

\newcommand{\be}{\begin{equation}}
\newcommand{\ee}{\end{equation}}
\newcommand{\bea}{\begin{eqnarray}}
\newcommand{\eea}{\end{eqnarray}}
\newcommand{\swifour}{Swi4}

\begin{document}

\title{Non-equilibrium dynamics of gene expression and the Jarzynski equality}
\author{Johannes Berg}
\address{ Institut f\"ur Theoretische Physik,
Universit\"at zu K\"oln\\
Z\"ulpicher Stra{\ss}e 77,
50937 K\"oln,
Germany}

\date{\today}

\begin{abstract} 
\noindent 
In order to express specific genes at the right time, the
transcription of genes is regulated by the presence and absence of
transcription factor molecules. With transcription factor
concentrations undergoing constant changes, gene transcription takes
place out of equilibrium. In this paper we discuss a simple mapping
between dynamic models of gene expression and stochastic systems
driven out of equilibrium. Using this mapping, results of
nonequilibrium statistical mechanics such as the Jarzynski equality
and the fluctuation theorem are demonstrated for gene expression
dynamics. Applications of this approach include the determination of
regulatory interactions between genes from experimental gene
expression data.
\end{abstract}
\pacs{87.16.Yc 
87.10.Mn 
87.16.dj 
}

\maketitle
\noindent
Cellular dynamics is based on the expression of specific genes at
specific times. The control over gene expression is a crucial feature
of nearly all forms of life, as it allows an organism to respond to
changing external and internal conditions.
With perfect regulatory control, only the DNA of those genes whose products are
required at a given instant would be transcribed to m(essenger)RNA
molecules. These mRNA molecules are in turn translated to
proteins. For example, enzymes to break down nutrients are produced
only when nutrients are present, or repair proteins are assembled to
respond to DNA damage.

To initiate the transcription of a gene, specific molecules, called
transcription factors, locate and bind to DNA near the starting site
of a gene. These molecules attract and activate an enzyme which reads
off DNA, producing an RNA chain molecule according to the DNA
template. Transcription factor molecules are themselves proteins and
thus subject to regulatory control, through other transcription
factors, or through themselves. As a result, mRNA and protein
concentrations of different genes may have highly non-trivial
interdependencies. A prominent example is the spatial-temporal
evolution of protein concentrations in the early stages of embryonic
development, leading to the formation of the body plan of an
organism~\cite{Davidson_book:2001}.

Despite the need for stringent control, gene regulation is an
inherently noisy process~\cite{McAdamsArkin:1997}. At the level of
single cells, only few molecules are involved, with single events
potentially having a large impact~\cite{Paulsson:2004}.

\begin{figure}[tbh]
\includegraphics[width=.49\textwidth]{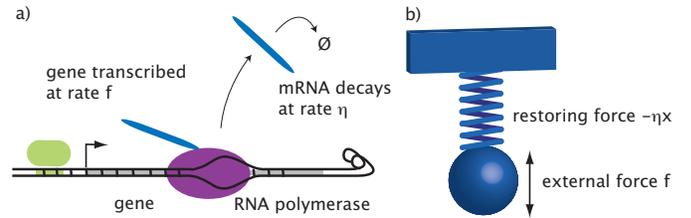}
\caption{\label{fig:mapping} {\bf Transcription and mRNA decay.}
a) Transcription of a gene is controlled by the binding of transcription
factors (left, shown in green) to the regulatory region of a gene. Transcription of a
gene leads to the production of mRNA molecules at some rate $f$. mRNA
molecules decay at a rate $\eta$ per molecule.
b) The resulting dynamics of mRNA concentration $x$ can be mapped onto
an harmonic oscillator subject to a restoring force $-\eta x$ and an
external force $f$ driving the system out of equilibrium.  
}
\end{figure} 

In this paper, the dynamics of mRNA concentrations in synchronized
cell populations is studied. The simplest model for the concentration
$x(t)$ of a given mRNA
is~\cite{Monodetal:1952,Ozbudaketal:2004,ChenEnglandShakhnovich:2004}
\be
\label{langevin1}
\partial_t x = -\eta x + f + \sqrt{D} \, \xi(t) \ , 
\ee 
where $\eta$ is the decay constant of the mRNA molecule and $f$ is the
average rate at which new molecules are produced by transcription of
the corresponding gene. The term $\xi(t)$ describes all other
processes, including changes in the transcription rate due to changing
transcription factor concentrations. Their influence has been modeled by a
random uncorrelated variable with mean zero and covariance $\langle
\xi(t) \xi(t')\rangle=\delta(t-t')$~\cite{Ozbudaketal:2004,ChenEnglandShakhnovich:2004}. Equation
(\ref{langevin1}) is well-known as the Langevin-equation of an
Ornstein-Uhlenbeck process describing the motion of an overdamped particle with
position $x$ in a quadratic potential $V(x)=(\eta x -f)^2/(2
\eta)$~\cite{vanKampenbook}. A thermal bath with inverse temperature
$\beta=2/D$ given by the Einstein relation exerts a random force
leading to an equilibrium solution $P_{\text{eq}}(x) \sim \exp\{- \beta V(x) \}$, see Fig.~\ref{fig:mapping}.

We probe this equilibrium scenario using experimental
measurements~\cite{Spellmanetal:1998} of expression levels of all
yeast genes taken at discrete intervals $\Delta_t$~\footnote{Expression levels
  give the amount of mRNA (converted to complementary DNA and relative
  to a reference sample) hybridised to a short strand of DNA on a
  so-called microarray chip~\cite{CarlonHeim:2006}. In the linear regime of
  hybridisation, expression levels are linear function of
  concentration. The data~\cite{Spellmanetal:1998} used here consists
  of $3$ sets of measurements  (termed alpha, cdc15, cdc28
  in~\cite{Spellmanetal:1998}) taken at intervals of $7$ to
$20$ minutes. A total of $59$ genomewide measurements were considered.}. 
In order to allow comparison across genes,  
we rescale the expression levels $x$ of each gene using $q =
\sqrt{2/(D \eta)}(\eta x-f)$ so the distribution of $q$ in equilibrium 
is $P(q) \sim \exp\{-q^2/2\}$. The parameters $\eta,f,D$ for each gene
were determined by maximizing the likelihood ${\mathcal P}_{\eta,f,D}({\bf x})$ of the expression
levels ${\bf x}\equiv\{x(t)\}$ with respect to the free parameters. 
The likelihood ${\mathcal P}_{\eta,f,D}({\bf x})= \prod_{t=1}^{T-1}
G_{\eta,f,D}(x_{t+\Delta}|x_t)$, where
$G_{\eta,f,D}(x_{t+\Delta}|x_t)=\frac{1}{\sqrt{2 \pi D \Delta}}
\exp{\left\{ -\frac{\Delta}{2D} (\partial_t x + \eta x_t - f)^2
  \right\} }$ is given in terms of the short-term propagator of the
Langevin equation (\ref{langevin1}). Drift and diffusion under this 
propagator can be compared in detail with the experimentally measured 
expression levels~\cite{Berg:inprep}. 

\begin{figure}[tbh]
\includegraphics[width=.31\textwidth]{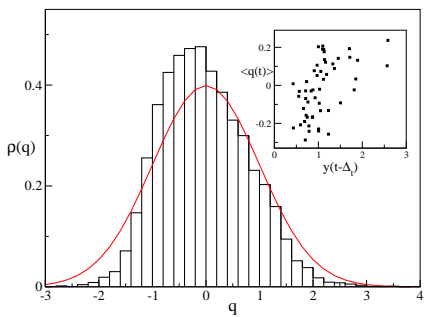}
\caption{\label{fig:q_dist} {\bf Empirical statistics of gene
    expression levels.} The set of (rescaled) expression levels of all
  ast genes at different times along the cell cycle has a distribution
roughly compatible with the equilibrium distribution of the Langevin
equation~(\ref{langevin1}) (solid red line). Deviations at high and
low expression levels might in principle be due to non-linearities of
DNA hybridisation to probes. Inset: However, 
  the distribution of
  expression levels is not stationary, but changes with the expression
  level of transcription factors. Here
  the mean expression levels $\langle q(t) \rangle_{\text{\swifour}}$ of
  \swifour~target genes at a given time $t$ 
  are plotted against the expression level $y(t-\Delta_t)$ of
  their transcription factor \swifour~at the preceding measurement. 
  The mean expression level of target genes is positively correlated
  with the expression level of the transcription factor, which changes
  continuously over the cell cycle.  
}
\end{figure}

Figure~\ref{fig:q_dist} shows the distribution of rescaled expression
levels $q$ across all genes and times. While the observed distribution
$P(q)$ is roughly
compatible with the equilibrium Gaussian distribution, the
statistics of expression levels is not stationary. As an example, we
consider the set of target genes of a transcription factor called
\swifour~\footnote{\swifour~is the DNA-binding component of a
  transcriptional activator, which regulates genes required for DNA
  synthesis and repair, as well as genes specific to the late G1 phase
  of the cell cycle. The name stands for ``SWItching deficient''~\cite{swi4_genecard}. 
The canonical binding sequence for \swifour~is
  ``CRCGAAA'' where R stands for either G or
  A~\cite{ChenHataZhang:2004}. Genes containing at least one copy of
  this binding sequence within $500$ base pairs from the transcription
  initiation site were considered target genes of \swifour.}. The
average value $\langle q(t) \rangle_{\text{\swifour}}$ of the target
genes at different times varies over the experimental time course, and
these average values are correlated with the expression levels of the
transcription factor \swifour, see inset of Fig.~\ref{fig:q_dist}.

This result is not unexpected: mRNA and protein concentrations of
transcription factors \textit{change on the same timescales as the
  concentrations of products of other genes}. Rather than the rapid 
fluctuations of the stochastic term in the Langevin equation (\ref{langevin1}),
the effects of transcription factors on their targets is a driving
force with a dynamics on the same timescale as that of the target
genes. In consequence, mRNA concentrations are kept out of
equilibrium.

These observations call for a non-equilibrium approach to gene
expression dynamics, which is the subject of this Letter. The
non-equilibrium regime is characterized by changes in the statistics
of gene expression levels over time. These are correlated with the
expression levels of the corresponding transcription factors. We model the
dynamics of mRNA concentration by the driven Langevin equation 
\be
\label{langevin2}
\partial_t x = -\eta x + f(y) + \sqrt{D} \, \xi(t) \ , 
\ee 
with the transcription rate $f(y)$ depending on the concentration $y$
of a given transcription factor at time $t$. This equation can easily
be generalized to describe the effects of several transcription
factors. The stochastic term $\xi(t)$ characterizes all
processes not yet described by $f(y,\ldots)$. In this sense,
(\ref{langevin2}) serves as a first starting point towards an
increasingly deterministic description of mRNA dynamics. In the
following, we will neglect post-transcriptional regulation and take
the mRNA expression level of a transcription factor as a proxy for its
protein concentration~\cite{Khaninetal:2006}. 

The equation of motion for the mRNA concentration (\ref{langevin2})
describes an overdamped harmonic oscillator subject to an external
force $f(y)$. Thus the dynamics of transcription factor concentration
$y(t)$ results in a time-dependent external force $f(t)\equiv
f(y(t))$. In the picture of a particle moving in a quadratic
potential, $V(x,t)=(\eta x -f(t))^2/(2 \eta)$ now is a time-dependent
potential whose origin changes with time. With each change of the
external force $\Delta f_t\equiv f_t-f_{t-1}$, with each change in the
potential, work is performed on the system. The total work performed
by the external force $f(t)$ between initial and final point of the
time course is denoted $W=\sum_{t_i}^{t_f} \Delta W_t$, with $\Delta
W= \left( \partial V/\partial f \right)_x \Delta f = - (\eta x-f)/\eta
\; \Delta f$.

The work $W$ quantifies the coupling of changes in the transcription
factor concentration to the mRNA concentration of a target gene and
serves as the central measure of the non-equilibrium approach. To
evaluate this quantity, we determine $f(y)$ within a simple model of
transcriptional activation: the probability of a transcription factor
being bound at a given binding site in the regulatory region of a
target gene depends on its concentration $y$, binding energy $\epsilon$, and
the free energy ${\cal F}$ of the transcription factor in solution or bound
elsewhere~\cite{GerlandMorozHwa:2002}. This model gives
\be
\label{michaelis-menten}
f(y)=f_0+ \frac{\delta \ y e^{-\epsilon/(kT)}}{y
  e^{-\epsilon/(kT)}+e^{-{\cal F}/(kT)}} \ , 
\ee 
assuming the transcription rate to depend linearly on the probability
that the binding site is occupied at a given time. $f_0$ is a basal
transcription rate in the absence of transcription factors and
$\delta$ quantifies the change of the transcription rate due to
transcription factor binding. The
functional form~(\ref{michaelis-menten}) is the celebrated
Michaelis-Menten kinetics, first studied in the context of enzymatic
reactions nearly a century ago~\cite{MichaelisMenten:1913} and used
widely in transcription modelling~\cite{Alonbook:2007}. The free
parameters of the model (\ref{michaelis-menten}) are inferred for
each gene from its mRNA concentration trajectory as above. 

\begin{figure}[tbh!]
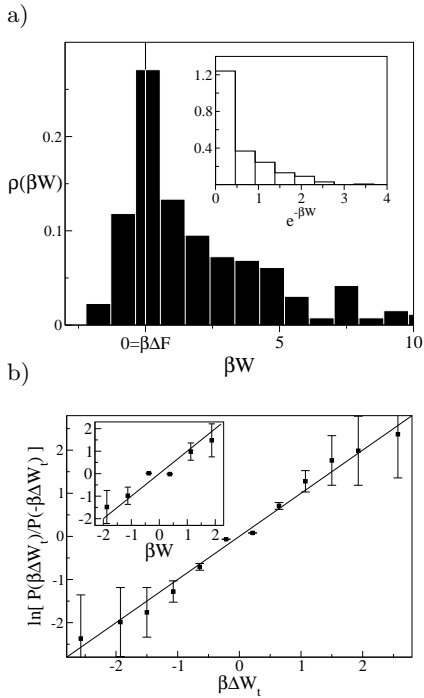

\flushleft{a)}
\vspace{1ex}

\includegraphics[height=.25\textwidth]{work_entrodisttk8_4.eps}
\vspace{-.5cm}
\flushleft{b)}
\vspace{2.5ex}

\hspace{1ex}
\includegraphics[height=.21\textwidth]{fluctthrm5locglob.eps}
\caption{\label{fig:jarzynski}{\bf The Jarzynski equality 
for gene expression.} 
a) The target genes of transcription factor 
\swifour~show a broad distribution of work $\beta W$ performed by changes in
\swifour~expression levels, with $\langle W \rangle > \Delta F=0$. 
Inset: The distribution of $\exp\{-\beta W\}$ has a mean of $0.96 \pm
0.33$ compatible with the Jarzynski equality. 
b) A \textit{detailed} relationship links the probabilities of paths with
positive and negative work performed, see text. The main figure shows 
the relationship for work $\Delta W_t$ performed between individual
timesteps, the inset shows the same relationship for the overall work 
$W$ performed over the full time course.
}
\end{figure}

Fig.~\ref{fig:jarzynski}a) shows, for different targets of the
transcription factor \swifour, the distribution of work $W$ performed
by changes in the \swifour~expression level over the time course. The
free energy $F$ of the equilibrium distribution of $x$, given by $\exp\{ -
\beta F\} = \int \, dx \exp\{ -\beta V(x)\} =\sqrt{\pi D /\eta}$,
does not change with $f$, since changes in the force $f$ only shift
the origin of the potential $V(x)$. The distribution of work for the
different genes obeys $\langle W \rangle \geq \Delta F=0$ as required
by the second law of thermodynamics. However, a small number of
trajectories has $W< \Delta F$.

A remarkable equality derived by C. Jarzynski~\cite{Jarzynski:1997}
links the work performed on the system averaged over many realizations
of the external forcing time course with the associated change in free energy,
\be
\label{jarzynski}
\langle \exp\{-\beta W\} \rangle = \exp\{-\beta \Delta F\} \ .  \ee
For a single trajectory of the system driven out of equilibrium by the
external force, $W$ is a random number depending on microscopic
details. According to the Jarzynski equality, however, the average of
$\exp\{ -\beta W\}$ over all trajectories equals $\exp\{-\beta \Delta
F\}$.  Its use in chemical reaction networks has been described
theoretically in~\cite{SchmiedlSeifert:2007}.

In a living organism, a specific time course of transcription factor
concentration is hard to repeat many times in order to perform an
average over trajectories. However, many target genes respond to the
time course of the transcription factor, and each target has a 
$W$ that is a random number which depends on the detailed
trajectory, but has a mean of $\exp\{ - \beta W\}$ equal to
$\exp\{-\beta \Delta F\}=1$. The inset of Fig.~\ref{fig:jarzynski}a)
shows the distribution of $\exp\{-\beta W\}$ across the target genes of
\swifour. It displays a broad distribution with mean and standard error
$0.96 \pm 0.33$ in agreement with the Jarzynski 
equality~(\ref{jarzynski})~\footnote{
The Jarzynski equality holds for initial conditions drawn from the
initial 
equilibrium configuration. A simple correction for initial
configuration not being drawn from the equilibrium distribution  
$\left(P_{\text{eq}}(q)/P_{\text{empirical}}(q)\right)$ is used here. }.

An even stronger statement holds, from which the Jarzynski
equality follows. Fig.~\ref{fig:jarzynski}b) shows the probabilities of
positive and negative work $P(W)$ and $P(-W)$ to be linked by a
\textit{detailed fluctuation theorem}~\cite{GallavottiCohen:1995,Crooks1999}
\be 
\label{detailed_fluct}
P(\beta W-\beta \Delta F=\beta w)/P(\beta W-\beta \Delta F=-\beta w) = \exp\{\beta w \} \ , 
\ee 
which shows how trajectories with work \textit{less} than the change
in free energy are exponentially less likely than those with work
performed in excess of the free energy change. This relationship can
be derived for generic time courses involving shifts of the origin of a
quadratic potential~\cite{Baiesietal:2006}. Thus the result that a detailed
fluctuation theorem holds for the work performed by the changing
transcription factor concentration serves as evidence for the linear
equation of motion~(\ref{langevin2}).

So far, we have focused on the statistics of mRNA concentration
trajectories given the parameters of stochastic models
like~(\ref{langevin2}). The reverse question, namely, what information
on transcription regulation can be extracted from experimentally
measured expression levels is an important question in systems biology
and bioinformatics~\cite{BussemakerLiSiggia:2001,Basso.etal:2005,Friedman:2004,BarJoseph:2004}. Some simple
attributes are already inherent in the observations of non-equilibrium
behaviour.  For instance, from the example in Fig.~\ref{fig:q_dist}
one can deduce that the transcription factor \swifour~acts as an enhancer
of transcription, rather than a repressor, since the average
expression level of its targets increases with expression level of
\swifour. Similarly, the targets of a transcription factor can be
determined from the inferred relationship $f(y)$ between the
expression levels of a transcription factor and that of a (potential)
target gene. This ``reverse engineering'' of regulatory interactions
is particularly relevant for transcription factors with
ill-characterized binding sequence, and for factors which do not bind
directly to regulatory DNA (so-called co-factors). For all
genes we compute the range of values of $f(y)$ over the range of $y$. Genes with a
large response $|f(y_{\max})-f(y_{\min})|$ to changing
transcription factor expression levels are presumed target genes. The
top ten targets of \swifour~predicted in this way are listed in
Table~\ref{table:swi4targets}. We test these predictions by searching
the regulatory regions of the predicted targets for copies of the
binding sequence~\cite{endnote32}. In all but one of the predicted
targets one finds at least one \swifour~binding site. Furthermore, $8$ of
the $10$ predictions have been previously found
experimentally~\cite{yeastract}. A more detailed account will be
published elsewhere~\cite{Berg:inprep}.

\begin{table}
\begin{tabular}{|l|l|l|}
\hline
CDC9 & 1 & \ding{55}  \\ 
RNR1 & 1 & \checkmark \\ 
YG3N & 1 & \checkmark \\ 
CRH1 & 1 & \checkmark \\ 
YIO1 & 1 & \checkmark \\ 
\hline
\end{tabular}
\hspace{.05\textwidth}
\begin{tabular}{|l|l|l|}
\hline
RAD27& 1 & \checkmark \\ 
PRY2 & 3 & \checkmark \\ 
CSI2 & 4 & \checkmark \\ 
PMS5 & 2 & \ding{55}  \\ 
CDC21  &0 & \checkmark\\ 
\hline
\end{tabular}
\caption{\label{table:swi4targets} 
  \small {\bf Predicted transcription factor target genes.}  The top ten predicted
    target genes of transcription factor \swifour~are listed along with the
    number of \swifour~binding sites in the regulatory regions of those
    genes~\cite{endnote32}. Check marks indicate existing experimental evidence for a direct
    regulatory interaction~\cite{yeastract}. About $3\%$ of
    the yeast genes have such direct evidence for regulation by \swifour. 
}
\end{table}

In summary, we have shown how regulatory interactions generate
correlations between expression levels of transcription factors and
their target genes. A simple mapping to a driven harmonic oscillator
depicts the transcription factor concentrations as an external force,
which drives the expression levels of target genes out of
equilibrium. Central quantity of this approach is the work performed
by the external force. Such dynamic observables provide a more
detailed fingerprint of the complex biophysical machinery behind gene
expression than heuristic measures like correlation coefficients.

It turns out that the work performed by the external force is of the
same order of magnitude as the temperature of the heat bath describing
stochastic effects, so $|\beta W| \sim 1$. Macroscopic systems
generally have $|\beta W| \gg 1$. As a result, experimental
observation of the fluctuations at the centre of the Jarzynski
equality and related theorems~\cite{Seifert:2007} has been limited to
the mechanical properties of
biomolecules~\cite{Liphardtetal:2002,HummerSzabo:2001} and colloidal
systems~\cite{Blickleetal:2006}. The correlated dynamics and complex
responses of gene expression offer a proving ground for stochastic
thermodynamics. Temporal data on other types of molecules apart mRNA
will lead to new challenges in the non-equilibrium dynamics of genetic
regulation.
\enlargethispage{1\baselineskip}

\begin{acknowledgments}
Funding from the DFG is acknowledged under 
grant BE 2478/2-1 and SFB 680. This research was supported in part by 
the National Science Foundation under Grant No. PHY05-51164.
\end{acknowledgments}


\end{document}